\documentclass[fleqn]{article}
\usepackage{graphics}
\begin{document}
\newcommand{\tab}{\hspace{5mm}}

\newcommand{\keywords}{quantum mechanics, Maxwell-Hertz electromagnetic theory,
detected signal} 
\newcommand{\PACS}{34.10.+x, 03.50.Kk, 84.47.+w, }

\newcommand{\email}{\tt vesely@itba.mi.cnr.it} 

\title{The Franck-Hertz experiment re-considered}
\author{S.\ L.\ Vesely$^{1}$, A.\ A.\ Vesely} 

\newcommand{\address}
  {$^{1}$I.T.B., C.N.R., via Fratelli Cervi 93, I-20090 Segrate(MI)
   \\ \hspace*{0.5mm} Italy \\ 
   }

\maketitle

{\small
\noindent \address
\par
\par
\noindent email: \email
}
\par

{\small
 \par
\par
\noindent{\bf PACS:} \PACS 
}

\begin{abstract}
\noindent The celebrated Franck-Hertz experiment is reinterpreted by analogy 
with the \textit{Glimmentladung} experiment, formerly performed by 
Heinrich Hertz.
\end{abstract}

\section*{Introduction}

The experiment of James Franck and Gustav Hertz in 1913 is considered 
crucial in quantum theory. In facts, it is said to bring evidence 
of the existence of discrete energy levels in the matter in the 
gas state. In 1949, Heisenberg still deemed it a fundamental 
experiment \cite{heisenberg1949}. However, that experiment 
and its successive improvements have been interpreted in the 
framework of the atomic model by Bohr and Sommerfeld. Those models 
definitively fell on the anomalous Zeeman effect, which the Heisenberg 
theory explained. At this point, we address to this specific 
issue a question that Schr\"{o}dinger put in general \cite{schro1983}: \textit{Do we mean this 
experiment to validate the} Kopenhagener Geist \textit{of quantum 
theory or else do we just admit that the interpretation of its 
experimental result is surviving the model that inspired it?} For 
sure, the atomic model played a key role in Heisenberg formulation. 
However, the Franck-Hertz experiment itself is not a
 \textit{Gedankenexperiment}. 
In this sense, it is not necessarily bound to interpretations 
leading to discrete energy levels.

Nowadays, there is a multitude of experiments that go under the 
Franck-Hertz suit. Even commercial devices exist, projected, 
built, and marketed for the sole purpose of performing the Franck-Hertz 
experiment. In this paper we present two cardinal experiments 
of those, developed by Goucher and Einsporn, respectively. They 
were performed at a time when the model to validate, the Bohr's 
atom, was still being tested. Those experiments, besides being 
interesting by themselves, abound of insights, more than enough 
with respect to the needs of the model. Nevertheless, they are 
not bullet-proof on critical points, as we'll see.

The Bohr's models and their subsequent transformations may work 
until they provide a way to grasp peculiar traits of the underlying 
natural phenomenon and successfully interpret experimental results. 
Our point is that, right now, the quantum theoretical background 
does not enhance the comprehension of the Franck-Hertz experiment. 
For comparison, we propose a rather classical electromagnetic 
interpretation of the phenomenon, based on a hertzian (after 
H. Hertz) kind of model-design.

We present historical facts, without getting much involved into 
historical questions. Rather, we extrapolate the hertzian train 
of thought enlightening it with latter-day concepts and knowledge.

\section*{On cathode rays nature}

The discovery of cathode rays was followed by a diatribe about 
their nature. In Poincar\'{e}'s words \cite{poinca1896},
``\textit{We see [...] on cathode 
rays the continuation of that same quarrel that took place during 
the Restoration on the light beams}.'' The quarrel was the same. 
However, about phenomena associated with cathode rays, the theories 
that contended for the best explanation were different. Rather 
than the mechanical theories by Newton and Huygens, the clash 
involved two electromagnetic theories: one by Lorentz and one 
by Maxwell.

As Poincar\'{e} aimed at a mechanistic reduction of electromagnetism, 
distinguishing between that kind of theories was not much relevant 
to him. On one hand electric phenomenology cannot leave out of 
consideration the matter constituting the source or the probe. 
To that extent, electricity and magnetism must be material phenomena. 
On the other hand, there are aspects in electricity that we are 
interested in and that have no mechanical correspondent. That's 
why no mechanical interpretation of electromagnetism is satisfactory. 
In other words, the rheologic behaviour of a viscoelastic fluid 
will never fully explain cathode rays propagation in a low pressure 
gas, nor radio waves at atmospheric pressure or in empty ether. 
As a concession to historical topics, it is interesting to observe 
that while theorists were still considering the possibility of 
mechanistic interpretations, experimentalists already agreed 
on the electric nature of cathode rays.

Those two electromagnetic theories, Lorentz's and Maxwell's, are 
different from each other \cite{roche1998}. Let's consider the latter first.
Maxwell's purpose 
is to frame Faraday's concepts and experiences within a mathematical 
model. He attributes an essential significance to the shielded 
cage experience. That is, he assumes the electrostatic charge 
to be a quantity with null algebraic sum. (That assumption is 
known as complete electrostatic induction). Thus, as the total 
free charge is invariably zero, hypothesising no mathematical 
limitation on the absolute length of dipoles, Maxwell formally 
handles the electric displacement field \textbf{D} by analogy with 
magnetic induction \textbf{B} in Farady's Law. In facts, he also refers 
to field \textbf{D} as electric induction \cite{maxw1954}.
This concept allows him to add 
a (supposedly small) displacement current term to Amp\`{e}re's 
equation. Such a formal amendment accounts for electromagnetic 
field propagation in the empty ether, and therefore it is not 
in contrast with Faraday's own concepts about fields. At those 
times, the ether was considered as a mechanical support for the 
propagation. It was the concept of propagation that called for 
the ether hypothesis: Maxwell's electromagnetic equations by 
themselves contain no dynamic variable.

Now about Lorentz's theory. He introduces a force to account 
for the mechanic effect of these fields on the matter. His expression 
for the ponderomotive force is inferred from electromechanical 
applications. However, Lorentz has to further state that electricity 
is inseparably coupled with elementary particles, the electrons. 
They are subject to a current continuity equation. The Maxwell's 
equations are carried over as-is into the new theory, although 
stressing their derivation from potentials. Hence, Lorentz reinterprets 
electrostatic induction, by attributing a mechanical mobility 
to the substantial electrical charge, and by giving conditions 
for its static equilibrium. Piling up of charges is not forbidden 
any more. Since Maxwell's mathematical field theory assumes zero 
total charge (a statement quite different from charge conservation) 
Lorentz's revision contrasts a postulate of the host mathematical 
theory \cite{poinca1901}. Really, 
that only affects further developments of the underlying mathematics.

As for a physical interpretation, we remark that each one 
of both scientists formally introduces a fictitious quantity 
in order to mathematically handle electric phenomena. The electric 
displacement the former, the electron the latter. For that reason, 
their theories won't support mechanical explanations.

Apparently, Heinrich Hertz is the last relevant experimentalist 
who opposes against Lorentz's theory. Indeed, he accepts the 
original Maxwell's equations and shows that his own experiments 
with cathode rays prove them. However, his detractors find Lorentz's 
theory to be conceptually simpler, and they interpret H. Hertz's 
observations after assuming the physical existence of electrons.

We repropose in an actual vein the interpretation of his experiment 
on the discharge in rarefied gases, because H. Hertz himself 
step by step builds a model for the Maxwell-Hertz theory. For 
an historical account of his writing and its impact, we redirect 
to J. Z. Buchwald \cite{buchwald1}.
 A newer publication by the same Author \cite{buchwald2} may give further 
insights.

\section*{Luminescence effects in gases as elicited by electric currents}

The paper by H. Hertz where cathode rays are considered as electromagnetic 
radiation, in accordance with Maxwell's theory is ``Versuche \"{u}ber 
die Glimmentladung.'' \cite{hertz1883}

It gives a clear account 
of experimental facts.

Our purpose in the following is to report the essential scheme 
of that paper with some comments.

H. Hertz asked himself three questions.

1)\tab 
Is an electric discharge through low-pressure gases continuous 
or discontinuous?

Let's explain this question. A gas, at a pressure ranging between 
1.5 and 0.01 mm Hg, is contained in a tube inside which two electrodes 
are sealed. These either consist of a pure high-melting metal 
or of an oxide coated one. Outside, they are connected through 
leading conductors and resistors to a battery or some other source 
of electric power. Experimentally, one shows that, even if the 
metallic circuit is not closed, current can flow through it. 
If the tube is thought of as a condenser filled with a dielectric 
mean, the flow of current is explained by Maxwell's equations, 
but the displacement current must be a function of time. On the 
other hand, if the flow of current is constant in time, then 
the tube operates in a non-linear manner. In such case, Maxwell's 
linear equations don't apply to its functioning.

H. Hertz answered question (1) on one hand by looking for alternate 
current (a.c.) components in the circuit, with different arrangements. 
On the other hand, he measured the Glimmlicht stroboscopically. 
That is a modification of the dielectric on the cathode side 
of the tube, which is observed during the time when current is 
flowing, and is similar in appearance to the glowing air before 
or during a spark breakdown of electrostatic generators. The 
Townsend discharge alone is a stable phenomenon up to the boosting 
power, where sparks shoot out. In a similar way, a tube containing 
a low pressure gas slowly discharges a battery, provided that 
it lights without breaking into pieces when the switch is closed, 
and that the power supplied is less than the one that starts 
and maintains the electric arc.

By hypothesis, a battery should supply a constant continuous 
current. However, the first researchers who experimented with 
batteries happened to draw an exceeding amount of power with 
respect to what the battery could deliver. Hence, they observed 
discontinuous glares, similar to Duddel's arc. H. Hertz increased 
the number of battery elements up to an open circuit potential 
of about 1.8 kV, and verified that the field-emission tube would 
not sustain possible oscillations of the circuit. In other words, 
he verified the absence of modulations at acoustic frequencies 
and up to intermediate frequencies.

This limit is imposed to him by the necessity of \textit{transducing} 
the electrical signal directly, i.e. without superheterodyne \cite{ravalico1943},
 into mechanical vibration. Neither moving-coil 
or needle electrical instruments, nor common microphones are 
loads matching electrical vibrations directly at frequencies 
higher than those considered by H. Hertz.

Because of the limits imposed by measuring devices, observable 
frequencies might arise from emf modulation, as mentioned above, 
and also from arcing drop of potential for the tube (this is 
the modulation technique used by ancient radiotelegraph Poulsen 
arc transmitters.) Receiving audio frequencies through the ether 
would have required an exceedingly long antenna.

H. Hertz concluded that, under his experimental conditions, there 
was no evidence of discontinuity, i.e. pulsation. That conclusion 
is enforced by current use of diodes as rectifiers. Strictly 
speaking, that means that Maxwell's equations do not account 
for continuous current through the tube, nor for cathode rays, 
nor for the bluish light wrapped around the cathode.

In facts, when we say that the bluish light or the Glimmlicht, 
are electromagnetic radiations, we are extending the applicability 
of Maxwell's theory significantly beyond intermediate frequencies. 
That is, beyond the zone that H. Hertz experimentally proved 
free from discontinuous currents.

Is that extension correct?

That extension invalidates the linear circuits theory: we know 
that no metal circuit conducts violet light. Furthermore, we 
know that many metals show resonance lines in the violet, that 
is, they emit without antenna. However, if the extension of the 
frequency range in Maxwell's theory is acceptable, the corresponding 
discontinuity of current might be identified with Schottky's 
noise. In that sense, the definition of noise looses somewhat its 
absolute meaning. The \textit{noise} becomes that part of the signal 
that, for whatever reason, we don't interpret.

2)\tab 
Do cathode rays follow the electric lines of force?

Let's put one consideration beforehand. If H. Hertz had a standard 
source of cathode rays at his disposal, he would already have 
mentioned them in the title of this paper. That's what he did 
in 1892 with the paper \textit{\H{U}ber den Durchgang der Kathodenstrahlen 
durch d\"{u}nne Metallschichten} \cite{hertz1892}. Rather, in 1883, the experimental 
difficulty was to establish what should be meant to be cathode 
rays, avoiding tautologies. In facts, those who discovered cathode 
rays did not go much further than ascribing a great deal of effects 
to that sole cause, without minding to distinguish between intrinsic 
properties and collateral effects.

Therefore, H. Hertz considered the tube as an element in the 
electrical circuit it belongs to, and assumed electromagnetic 
field equations to apply to it at steady conditions. The same 
hypothesis is being used today to avoid graphical methods when 
analysing electrical circuits: one draws the wiring diagram and 
replaces the diode symbol with its equivalent form. H. Hertz 
used linear analysis to interpolate inside the tube the measurements 
he took just outside it using a small magnet. He traced the lines 
of current of the static field, after Maxwell. He found 
that neither the cathode nor the anode luminescence follow those 
lines.

The result he found calls for two separate explanations: one 
for the current in the circuit and one for the cathode rays.

In primis, we have to interpret the lumped elements wiring diagram: 
the emf supplied is shared between the load resistance and the 
tube resistance at the operating conditions. The tube resistance 
is determined according to the static volt-ampere characteristic. 
In such a diagram, the figurative place where the flow tube of 
Maxwell's field lays is filled with the symbol of a lumped resistor, 
so in no place the current goes through the empty ether.

Then, there is the Glimmlicht. It is a collateral effect with 
respect to the use of the tube in the circuit. Indeed, thermoelectric 
valves that really took root make no use of it. Furthermore, 
the electric agent associated to the Glimmlicht in low pressure 
gases (and possibly to their enhanced conductivity) does not 
contribute to the electric current in the circuit, both visually 
and after Maxwell's theory, but dissipates all around. Is it 
possible that this agent, which is the active emission of a fed 
valve, behaves as an electromagnetic radiation in Maxwell's sense 
without being described by the circuit equations? Yes it is. 
The branch of electromagnetism devoted to it is known as radiotelegraphy. 
In the circuit, that emission accounts for small fluctuations, 
i.e. mathematically higher order terms. You may consider that 
as H. Hertz definitive answer, if you like to. However, it might 
not be deduced from this experiment, since a diode is no transmitter. As 
he himself acknowledged in 1894: ``In the beginning, I thought 
that electric motions were too harsh and too rough to be useful''\cite{hertz1993}.

3)\tab 
Do cathode rays exhibit electrostatic properties?

In the preceding question (2), H. Hertz ruled out the investigated 
electric agent to contribute to the mean circuit current. That 
does not mean that current and emission are not related to each 
other. In a much similar way, in question (3) H. Hertz asks himself 
whether cathode rays charge the matter they strike, since they 
consist of an electrically charged flow. He does not ask himself 
if cathode rays could be received, demodulated and detected as 
a faint current. Nor he puts up with electric wind.

Let's explain better the implications of the possible answers. 
In Lorentz's case, one conjectures that the neat negative charge 
radiates or, if you prefer a hydrodynamic analogy, that it slowly 
flows through space as time goes by. Then, charges may accumulate 
on dielectric media, according to the continuity equation, and 
the time integral measures the piled up ponderable or imponderable 
charges. In the other case, Maxwell's theory does not explain 
how to bring in any static electricity \cite{darrigol2000}.
 That's because electromagnetism formally meets the elder 
Coulomb's theory of \textit{static} ``action at a distance'' whilst 
it is rather obvious that contact, rubbing, and similar ill-theorised 
operations play a key role in electrification. Also recall that 
the elder theory ``at a distance'' for sure didn't mean displacements 
of material points subject to \textit{static applied forces} to take 
place instantaneously.

Nowadays we still say, about friction electrical machines, that 
the electricity of a sign is being conduced to ground, and we 
imagine that the other ``tank'', of less capacity, gets charged 
with the opposite sign. However, when both tanks have the same 
capacity, as e.g. in Nairne electrical machine, we consider charges 
of opposite sign being just separated, and we represent the pads 
charged at equal and opposed voltages. This non-mechanical duality 
of electric displacement doesn't change from the old to the new 
field theory.

Faraday, rather than reckoning the amount of electric charges 
already separated on the conductors pads, considered the polarisation 
of the interposed dielectric mean. In other words, he prevented 
the charges from arousing directly in the locations corresponding 
to the physical surfaces, by imposing continuity by spatial contiguity \cite{maxw61}.
 Thus far, we cannot deduce that, by the end of 
the process, something that wasn't there before has accumulated 
on the plates or in between them. That way, the results of the 
electrostatic experiments are preserved in the new field representation. 
Hence, it is the accumulation of net charge that makes the difference 
between the two theories, keeping in mind that Maxwell's equations 
provide for net charge as a boundary condition.

H. Hertz decided to consider the negative charges on the glass 
walls as if they were given steady conditions. He measured them 
after transient, on starting the first experiment of each day. 
He recorded the cathode ray emission by detecting its fluorescence. 
More precisely, he detected the fluorescent signal of the glass 
surface on the bottom of the tube, which was completely shielded 
from the electrodes.

His generator incorporated a Ruhmkorff coil, so current was \textit{alternate}. 
H. Hertz was forced to the choice of that generator, and therefore 
he had to check that Maxwell's term for the displacement current, 
whatever it physically means, didn't impair cathode rays. Basing 
on his own experiments, he excluded that the glass, besides fluorescing, 
also got charged. More precisely, he wrote that near to the cathode, 
upstream of the shielding, the glass envelope got at a negative 
potential in a durable way. He didn't relate that fact directly 
with cathode radiation. He also observed that, by applying an 
additional electrostatic field, the breakdown threshold was lowered, 
and neither that he considered to be a first order effect.

The missing detection of the alternate current has been puzzling 
the interpreters of his experiment subsequent to himself.

Heating the cathode was an innovative technique at the time. As H. Hertz
doesn't mention it, we presume he used no hot-cathode device in this experiment.

We are suggesting that H. Hertz's diode didn't detect 
alternate current because cold-cathode diodes do not rectify it \cite{easman1949}.

His negative answer to the third question is remarkable, but for 
the time being, we just like to pinpoint consequences of H. Hertz's 
(and J. Maxwell's) choice that are under everybody's nose.

The superposition principle applies to (homogeneous) solutions 
of wave equations, not to a flow of charge. For the same reason, 
frequency spectra analysis can characterise waves, not flow of 
charge. A phenomenology requiring an extensive development of 
the flow of charge theory would not have lead to telecommunication 
technology as we know it today.

We don't mean that hertzian experiments were aiming at telecommunications 
development, as the latter technology didn't exist as a project, 
before H. Hertz. His discovery is different: to whom who stops 
considering electromagnetic induction as just a property of the 
magnetic flow, which induces an emf in the relevant circuit \cite{slater1969},
 electromagnetic \textit{far-field} theory may appear as 
a new (linear) physical theory. Otherwise, it is just bare mathematics, 
without the multiplicity of electrical displays on the matter \cite{planck1899}.
 In his own words \cite{hertz1993}: ``It was the distance at which I could perceive 
the action, getting larger and larger, what aroused my utter 
astonishment. Until then I was used to see electric forces diminishing 
with Newton's law, i.e. rapidly vanishing as distance increased.''

\section*{The electric measurement of excitation and ionisation potential 
of gases}

Besides the so-called silent electric discharge, luminous emission 
from gases can be obtained by heating or with sparks. As techniques 
became finer and finer, an empirical rule for emission was stated 
as

\begin{eqnarray}
  1234 = {\rm wave length} \times {\rm voltage drop\ (nm V)}
\label{eq:A}
\end{eqnarray}

Indeed, without a systematic approach, the phenomenology looked 
more puzzling. On one hand, the experiments with enough gas pressure 
to make the Glimmlicht visible agreed that there should be a 
relation between the electrical parameters of the gas valves 
and its (cold) emission spectrum. This very fact was expressed 
by the \textit{simple relation} written above. On the other hand, 
the electrical functioning of the valves themselves satisfied 
no linear relationship.

Truly, electric ionisation experiments were sophisticated ones. 
Furthermore, at those times it was believed that natural laws, 
being concerned with material bodies as sensible objects, could 
eventually be expressed as dynamical laws. (That is, e.g., after, 
reducing phenomena from a thermodynamic description to a statistical 
mechanics formulation.)

As there is no easy mechanical way to lead a gas into emitting 
light, we may say that valves of the De Forest type, i.e. triodes, 
tetrodes, etc., were used because they were supposed to emit 
electronic particles endowed with mass \cite{hund1984}. 
The experiments were devised in order to accelerate those particles 
by applying an electrostatic potential, measuring the values 
at which the impulse would have been transferred to an atomic 
gas by means of inelastic conservative collisions. As it is still 
currently being taught, at the potentials where inelastic collisions 
take place an electron-bullet wouldn't have enough kinetic energy 
to hit the anode. At the same time, after a gas atom target has 
absorbed its impulse, it would transfer it on a higher energy
Coulombian/Newtonian orbit of its optical electron.

Those measurements were not quite mechanical, but they rather 
involved electric quantities -- the small variation in the continuous 
anode current. Nevertheless, researchers could readily imagine 
electrons escaping the cathode material toward the interspace. 
They imagined the electrons and/or the positive ions produced 
in the collisions would have been gathered on the anode. To wit, 
they thought of the electrical feeding essentially as a mean 
for boiling electrons off the cathode.

The researchers who carried out those experiments deemed them \textit{fundamental,} 
that is oriented toward understanding nature at a most basic 
level, in contrast to other contributions to electronics, more 
aimed at applications and technology.

That way, H. Hertz's basic idea that a cold-cathode diode fed 
with power from electric batteries behaves as an electrical lamp\footnote{See 
T. A. Edison 1883, for thermoionic type of lamps.}, and that 
its emission could, in first approximation, be described by Maxwell's 
equations, seemed to be missing an essential aspect of electrical 
phenomena.

Stemming from the lack of familiarity with bare electrical explanations, 
Bohr and Sommerfeld's atomic models aim at \textit{rationalising} relation 
(\ref{eq:A}) above, i.e. bringing it down to mechanics. The experiments, 
controlling temperature, purity and concentration of gases, and 
also polarisation and oxidation of electrodes, etceteras, aim 
at verifying the \textit{numerical correspondence} between the frequency 
values computed after law (\ref{eq:A}) of the model and the optical spectra 
recorded directly.

With respect to that verification, Franck and Einsporn expressed 
themselves with words like the following \cite{franck1920}.
 ``\textit{We ask ourselves why the lines 
corresponding to the transitions 1.5S -- mp1 (wave length 2656 
{\AA}) and 1.5S -- mp2 (wave length 2270 {\AA}) of mercury vapour, 
that should stay in some easily accessible ultraviolet zone, 
have never been observed up to date in neither emission nor absorption 
optical spectroscopy. In particular, our measurements suggest 
that the line corresponding to 1.5S -- mp2, that we clearly see 
on our graph as a sudden increase in current at about 5.43V, 
should be allowed}.''

Since 1926, the Franck-Hertz experiment has been dealt with using 
the mathematics of self-consistent field theory \cite{born1969}.
 If that is its explanation, 
the Franck-Hertz experiment ceases to be \textit{fundamental} in that 
it isolates a distinct natural phenomenon and lays it bare for 
comprehension by a physical theory. It may be considered a fundamental 
experiment for some other reason, such as, e.g., supporting the 
aforementioned mathematical theory. The point is that its conceptual 
binding to quantum mechanics becomes somewhat loose \cite{cassin1996}.
 It appears as if one could not deduce any failure of the probabilistic quantum 
theory from experimental results of experiments like this.

Now observe that quantum mechanics assumes a unique ``mechanical equivalent of 
electricity.'' A concept that Lord Kelvin traces back to Aepinus \cite{phil1902}.
 It is manifest in the Lorentz force, 
which represents a sort of electromechanical transduction. That 
law cannot be easily extended from atomic behaviour to the efficiency 
of a working engine. Anyway, if it held true for electrical circuits, 
then impedance matching would be an option even when designing 
high frequency transmission lines. On the opposite, experience 
suggests that a significant mismatch is not compatible with gain 
and high power levels already at higher audio frequencies.

From a modern point of view, the H. Hertz experiment described 
above can be seen as an unusual diode application. In facts, 
diodes are used as rectifiers more often than not. Likewise, 
experiments of the Franck-Hertz type can be considered unusual 
applications of triodes and tetrods. Are they fundamental only 
because they can be explained theoretically? Then, one may ask, 
why doesn't the underlying theory rely on widespread applications 
of those valves?

Consider Maxwell-Hertz theory. Although it is not founded on 
widespread applications of valves, in practice it has been supporting 
their implementation ever since. Perhaps unusual applications 
are more difficult to interpret within that theory simply because 
the experimental design is just not aimed at that.

\section*{F. S. Goucher and the Barkhausen-Kurz generator}

Goucher \cite{roy1916} has been one of the researchers who contributed methodological 
improvements to the original Franck-Hertz experiment. With respect 
to H. Hertz time, feeding, thermostatation, and vacuum technique 
have improved. The purity of the mercury certainly has a higher 
standard. Most importantly, there is one more electrode in the 
tube: a grid (that Goucher called \textit{gauze}). That's an important 
innovation, as the third electrode can modulate the electromagnetic 
response of the valve.

\begin{figure}
  \resizebox{8cm}{5cm}{\includegraphics{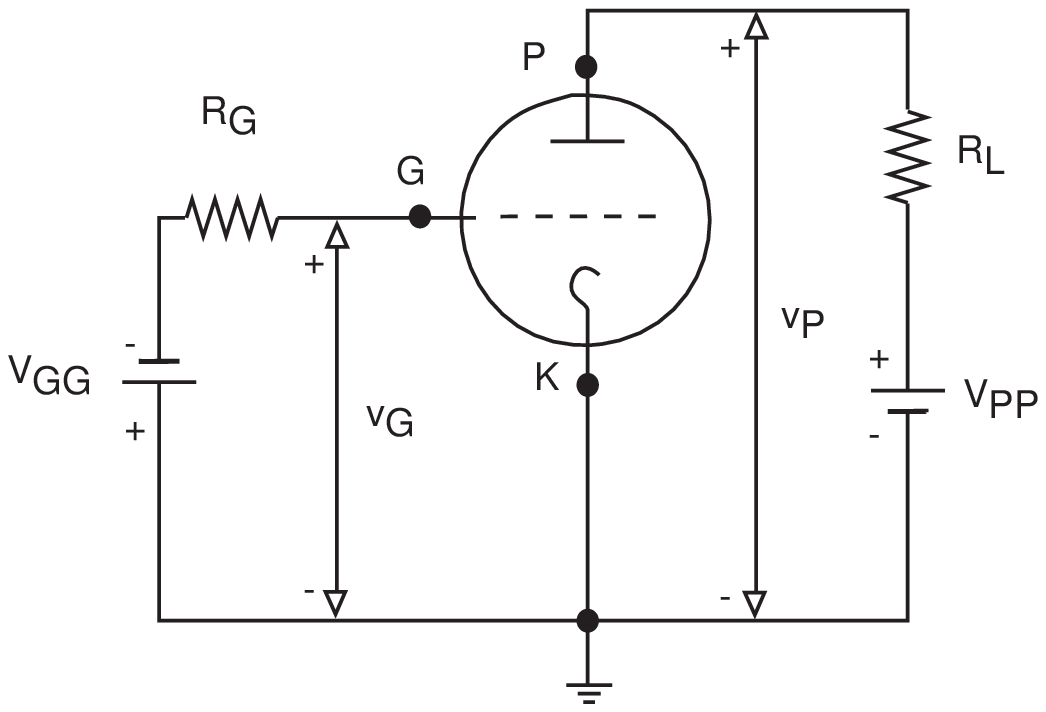}}
{\small\\
P is the plate\\
G is the grid\\
K is the cathode\\
$V_{GG}$ is the grid supply voltage\\
$R_G$ is the grid static resistance, if this electrode is drawing a current $i_G$\\
$V_{PP}$ is the plate-to-cathode voltage bias\\
$R_L$ is the load resistance\\
if $v_G$ is the input potential, $i_P$ is the output current\\
Cathode-heating circuitry is omitted
}
  \caption{A basic grounded-cathode circuit
}
  \label{fig_1a}
\end{figure}

Better on that concept. H. Hertz measured only one curve of current 
corresponding to potential drops across the diode, given load 
and battery. In the triode, the non-linear relation F(V, I) depends 
on two more parameters, as one may vary potential and current through 
the third electrode too.

In electronics, the superior versatility of that component leaded 
to the standardisation of a few families, depending on applications. 
Let's examine two usual applications of vacuum tubes.

As a first example, consider high fidelity amplification of weak 
electromagnetic signals at audio frequencies. We restrict to 
the essential, omitting even the transduction step into audible 
sound. Fig. \ref{fig_1a} shows the basic schema of a triode configured as 
a common-cathode wide band amplifier, with grid and plate supply 
batteries.

Amplification being untuned, the output signal, i.e. the plate 
current on an external resistive load, at the operating conditions 
is the linearly amplified input signal, i.e. the variation of 
the grid potential. Then, the circuit can be represented as a 
two-port network having a lumped output impedance in series with 
a generator, equal to the open circuit voltage.

\begin{figure}
  \resizebox{7cm}{4cm}{\includegraphics{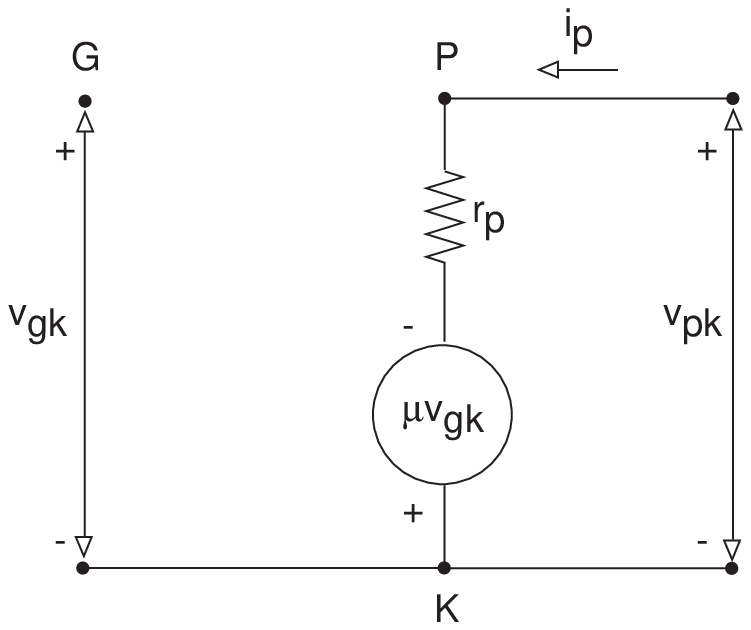}}
{\small\\
  $r_p  \equiv \left( {\frac{{\Delta v_P }}{{\Delta i_P }}} \right)_{V_G }$ is the plate resistance\\
      $\mu  \equiv  - \left( {\frac{{\Delta v_P }}{{\Delta v_G }}} \right)_{I_P }$ is the amplification factor\\
      $v_{gk}  = v_G  + V_{GG}  = \Delta v_G $ is the input voltage drop\\
      $v_{pk}  = v_P  + V_{PP}  = \Delta v_P $ is the output signal\\
      across load resistance $R_L$, i.e. $v_{pk}  =  - \frac{{\mu v_{kg} R_L }}{{R_L  + v_P }}$\\
      Configuration is grounded-cathode. Quiescent point\\
      values on a static plate characteristic\\
      $V_{GG} = const, I_P = I_P(V_{PP})$ are not stressed\\
}
  \caption{
Thevenin's linear model for the updated range over which
$\mu$ and $r_p$
are substantially constant.
}
  \label{fig_1b}
\end{figure}

 Fig. \ref{fig_1b} shows 
the schematic voltage-source model where supply batteries are no 
more included, as usual.

As a second example let's consider the generation of a carrier 
for radio broadcasting (without radiation step ``coupling'' to 
the ether channel). The basic scheme of Fig. \ref{fig_1a} for the triode 
may be kept valid if we substitute a parallel circuit \textbf{LC} 
for the resistive load \textbf{R}$_{\mathbf{L}}$. The new element plays the 
role of a passband filter at the carrier frequency. However, 
the simple scheme in Fig. \ref{fig_1b} doesn't apply any more, because now 
amplification is tuned. Furthemore, the input has to be obtained 
feeding back part of the output. If the feeding back keeps the 
correct input-output phase relationship an oscillation starts 
and builds up as the biased amplifier is matched at the correct 
frequency and tracks it. Technical improvements allow to get 
a steady carrier with low harmonic content, low levels of sideband 
noise and low thermal drift rate. At variance to the first example, 
the output is typically on/off. It is not linear, nor it can 
be obtained from linear amplification by a perturbative approach.

The quirk in using a triode as in Franck-Hertz experiments doesn't 
lay in any dissimilarity from the scheme in Fig. \ref{fig_1a}, but rather 
in its atypical circuitry, with respect to usual applications.

Really, Goucher used the third electrode to control the valve, 
but he didn't bother about ``tube characteristics.'' The reason 
is that he was interpreting his observations after hypothesising 
a slow flow of particles. By that hypothesis, the investigation 
became concerned with the distribution of velocities of those 
particles. The non-linearity of the characteristics was considered 
``experimentally uninfluential'' until it didn't increase the 
dispersion around the mean value of that distribution. 

Now, let's substitute those thermo-emitted particles accelerated 
via an applied d.p., which ionise the gas because of collisions, 
with the alternative picture. Let's say that, since the casing 
and contents behave as a triode, the signal distortion arising 
because of the non-linear characteristics explains the presence 
of an output without external input.

We show that Fig. 3 of Goucher's paper (which we number as 
Fig. \ref{fig_3}) is a characteristic curve of that valve, and that its 
non-linearity justifies the onset of persistent autogenous oscillations 
of his thermoelectronic tube at the expenses of the plate feeding 
power.

We like to start by telling about characteristics of vacuum tubes. 
As we said above, about H. Hertz experiment, if we consider a 
diode as a planar equipotential-surfaces condenser, then, according 
to electrostatics, the potential between the electrodes is a 
linear function of their distance, and to a good approximation 
there is no current flow. In disagreement with electrostatics, 
a thermoelectronic diode can conduct, and, although to a far 
lesser extent, also a cold-cathode diode can. The relation between 
the continuous current through it and the potential drop, its 
static characteristic, is non-linear.

\begin{figure}
  \resizebox{8cm}{4cm}{\includegraphics{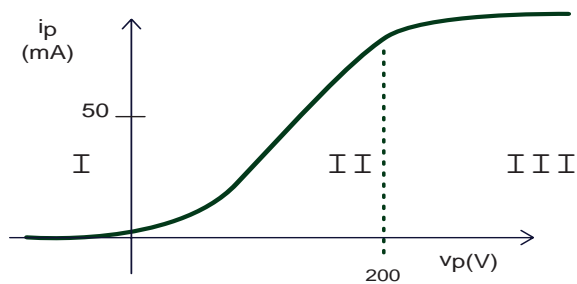}}
  \caption{Shape of a volt-ampere diod's characteristic for a high-vacuum tube
    with pure tungsten cathode at T $\simeq 2200\ $K 
    }
  \label{fig_2}
\end{figure}

As shown in Fig. \ref{fig_2}, qualitatively, we distinguish three ranges 
of a commercial thermoelectronic diode's characteristic. There 
is a lower saturation range (I) at negative potential, where 
the diode approximately doesn't conduct. There is an upper saturation 
range (III), where the maximal current essentially depends on 
the cathode's hot filament supply. Then, there is an intermediate 
range (II) where the relation between
\textbf{i}$_{\mathbf{P}}$ and \textbf{v}$_{\mathbf{P}}$ 
is considered to depend on the surface conditions of the cathode, 
on the way it's feeded, on the material of the electrodes, on 
their geometry, and on the residual gases in the tube. The trend 
may seem quadratic, but probably it is even more complicated.

In the following few paragraphs, we describe the curve for a 
vacuum tube having three electrodes.

We already mentioned that the grid control introduces a further 
parameterisation of every isothermal characteristic function. 
In facts, the common cathode triode may be considered as composed 
of two interlaced diodes: an input diode, consisting of the cathode 
and the grid, and an output one, consisting of the same cathode 
and the anode.

Consider the static parameters v$_{\mathbf{G}}$, v$_{\mathbf{P}}$, i$_{\mathbf{G}}$ and 
i$_{\mathbf{P}}$ of the triode, where \textbf{G} (for grid) indexes the values 
pertaining to the input diode while \textbf{P} (for plate) those of 
the output one. By a careful design, the amplification can be 
made quite linear, with an amplification factor between 10$^{2}$ 
and 10$^{4}$ at the clamps, in a certain range of the static parameters. 
In that range, the parameterisation of the static characteristic 
is a repetition of the same curve at different offsets. To wit, 
a given output characteristic (v$_{\mathbf{P}}$, i$_{\mathbf{P}}$) generates regularly 
repeated graphs for a set of fixed v$_{\mathbf{G}}$`s that differ from 
one another by a constant amount.

We have v$_{\mathbf{G}}$ = v$_{\mathbf{G}}$(i$_{\mathbf{G}}$),
and the input diode characteristic 
(v$_{\mathbf{G}}$, i$_{\mathbf{G}}$) is non-linear as a whole. That's why amplification 
without distortion typically implies i$_{\mathbf{G}}$\ensuremath{\sim}0, that 
is this diode is cutoff. In turn, this means that the grid's 
potential has to be negative w.r.t. the cathode. Hence, we see 
that the purpose of using a positive grid potential to accelerate 
electrons may become incompatible with a nicely linear response 
of the triode. This terminates the digression about tube's characteristics.

Back to the aforementioned Goucher's paper. The stated aim of 
Fig. \ref{fig_3} is to show that the triode being used is more suitable 
for measuring ionisation potentials than Pawlow's one. So we 
may compare two performances and check what's the technical improvement.

\begin{figure}
  \resizebox{8cm}{5cm}{\includegraphics{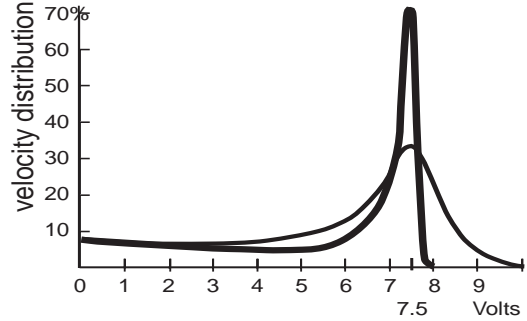}}\\
  {\small Tracing according to Goucher's Fig. 3 of \cite{roy1916}}
  \caption{
    Goucher's and Pawlow's experiments
    }
  \label{fig_3}
\end{figure}

Goucher's tube featured an indirectly heated pure metal cathode 
(Pt). The vacuum obtained from hydrogen was better than 0.005mm 
Hg (which is still about 10$^{3}$ times worse than high vacuum in 
1949.) Measurements were carried out maintaining the grid potential 
v$_{\mathbf{G}}$ at +7.5V, varying the static grid-plate decelerating 
potential v$_{\mathbf{GP}}$ between 0 and 10V point by point, and registering 
the corresponding plate electrical current by means of an auxiliary 
field electrometer when stationary conditions were attained.

In Fig. \ref{fig_3}, on the ordinate axis he plotted the calculated slope 
of the experimentally measured i$_{\mathbf{P}}$ as a velocity distribution. 
The calibration was the percentage of the graph area, between 
0\% and 70\%. On the abscissa, he plotted the values of v$_{\mathbf{GP}}$ 
between 0 and 10V.

Since v$_{\mathbf{PG}}$ = v$_{\mathbf{P}}$ -- v$_{\mathbf{G}}$, and v$_{\mathbf{G}}$ = 7.5V,\\
v$_{\mathbf{PG}}$ = 0 implies v$_{\mathbf{P}}$ = 7.5V\\
v$_{\mathbf{PG}}$ = -10 implies v$_{\mathbf{P}}$ = -2.5V.
\\
\\
On the other hand, we have

\begin{eqnarray}
\left( {\frac{{\partial {\rm i}_P }}{{\partial v_{PG} }}} \right)_{{\rm v}_G }  = \left( {\frac{{\partial {\rm i}_P }}{{\partial {\rm v}_P }}} \right)_{{\rm v}_G }  = {\rm g'}_P  = \frac{1}{{{\rm r'}_P }}
\label{eq:B}
\end{eqnarray}

Therefore, the graph he plotted may also show a plate static 
differential conductance as a function of the plate potential. 
The apex indicates that, if amplification is not linear, conductance 
differs from that one for small swings around the chosen working 
condition. We'll now argue that Goucher's project aimed at enhancing 
the non-linear functioning.

When the common cathode triode is used as a voltage amplifier, 
that is it exploits the linear characteristic range, it exhibits 
a high plate resistance, which contributes to the remarkable 
value of the linear gain factor $\mu \propto r_{P} $.

 That might hold for both Goucher 
and Pawlow tubes, up to about half of the abscissa range. Thereafter 
v$_{\mathbf{P}}$ tends toward 0, while the input diode is not cutoff. 
That implies \textbf{i}$_{\mathbf{G}}$ \texttt{>} 0: Goucher doesn't say it for 
this triode, but for the one on page 570 of his paper. This current 
flows in the input diode, generating a voltage drop, and hence 
it supplies to the output an additional voltage ``in phase''
with \textbf{v}$_{\mathbf{P}}$.

According to commercial data sheets, the positively polarised 
grid draws more and more current as the plate potential tends 
toward zero.In the same conditions, according to Fig. \ref{fig_3}, Goucher's 
tube conductance climbs showing a much narrower ``gaussian'' peak 
than Pawlow's. Moreover, it grows in spite of there being no 
external user requiring the current to increase. Finally, it 
abruptly hits the abscissa axis with a distinctively not horizontal 
slope. For the remaining 2V, only the slow descent of Pawlow's 
tube is reported in Fig. \ref{fig_3}. Perhaps Goucher got negative conductance?

As far as the area under the graph is concerned, a very steep 
tract with negative conductance wouldn't contribute much, and 
hence any correction bears no importance. However, for the sake 
of the triode electromagnetic response, to exhibit a negative 
conductance implies an instability of amplification. To be precise, 
for those bias conditions the triode performance looks like that 
of a generator. In facts, the Barkhausen-Kurz generator for U.H.F. 
is connected and polarised nearly the same way \cite{henney1950}.
 Obviously, an 
oscillator circuit usually contains also filters and provides 
for a matching at the desired frequency of oscillation. Goucher's 
circuit consisted of just the tube. That notwithstanding, according 
to this interpretation, it oscillated.

So, this is the big difference between the diodes that H. Hertz 
used to work with and these fine workmanship pure metal cathode 
triodes. The former couldn't generate any radiotelegraphic carrier. The 
latter typically resonate in certain frequency ranges. Indeed, 
between 1928 and 1929 commercial triodes and tetrods have been 
substituted with pentods for linear amplifiers, or they have 
been cascode configured.

As a further look inside the technological development trend 
of those experiments, we also report an improvement suggested 
by G. Hertz for noble gas, helium-neon or metal vapour, triodes: 
``\textit{The heated cathode [...], with a drop of barium oxide on 
its surface, is at about half a millimetre from the close-meshed 
grid. [...] The space between the grid and the plate is an enclosed 
metallic cavity, except for a small slit [...]}''\cite{hertz1924}. 

We conclude that Goucher's efforts were directed toward improving 
the grid-plate coupling of his tube, in order to increase its 
positive feedback and quality factor, by keeping losses as low 
as possible.

\section*{E. Einsporn and the voltage controlled oscillator}

Einsporn \cite{zsfp1921}, one of Franck's co-workers, considered a common-cathode 
configured \textit{thermoionic} tube of four electrodes.

He made two different series of measurements: in the first one, 
he measured mercury excitation potentials, using the tetrod as 
a triode. In the second series, he measured ionisation potentials, 
for comparison with Goucher, using the tetrod as a diode.

In the first series, the Author applied the ``accelerating'' positive 
potential between the cathode and the second grid, that is the 
one near to the plate. Now, when the first grid is earthed, the 
tetrod works as a triode. Furthermore, as the amplification gain 
is related to the ratio between the geometrical plate-to-grid 
and grid-to-cathode distances, we may consider it small, except 
at those working conditions where the input couples with the 
output.

During the experiments, the grid potential was set at a chosen 
value between 0 and 25V, whilst the plate one was held slightly 
positive w.r.t. the grid (max. 1V). The corresponding measure 
of current was taken. The whole procedure was to be repeated 
for each value of the grid potential.

\begin{figure}
  \resizebox{7cm}{12cm}{\includegraphics{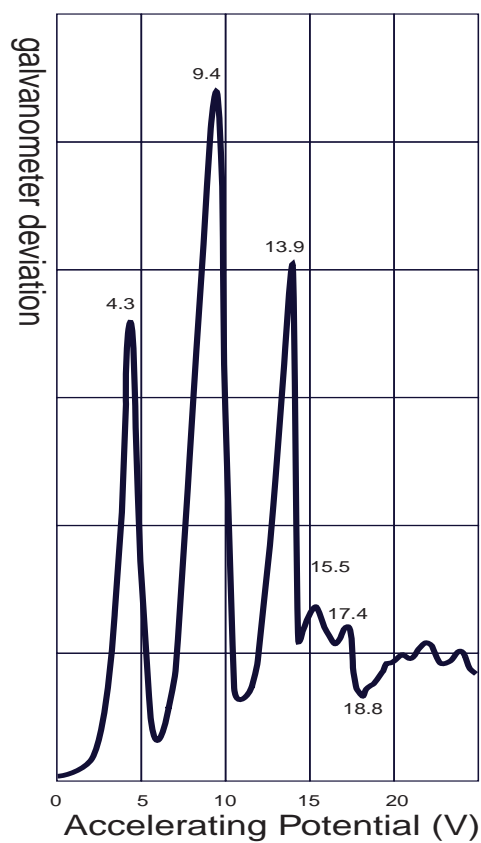}}\\
  \caption{
    Tracing of Fig. 3 according to Einsporn \cite{zsfp1921}
    }
  \label{fig_4}
\end{figure}

The curve reported in his Fig. 2 and redrawn in our Fig. \ref{fig_4} to interpolate 
his data for the plate current as a function of the grid potential, 
is not a mutual characteristic in the prevailing sense of this 
term. In facts, it consists of points from a set of data that 
we can regard as plate characteristics. Now, if the whole valve, 
i.e. outer casing, electrodes, wiring and gas, only amplified 
more or less linearly the static input potential in the considered 
biasing range, then the slope of that curve should exhibit no 
irregular climbing, or at least it shouldn't change sign. Instead, 
Einsporn himself calculated interpolations as if the curve had 
several narrow local maxima. These are well distinguished at 
the grid potential values A, 2A, 3A and their products with another 
potential B. As usual, he related those peaks to the UV spectrum 
of the gas inside the casing, by means of the simple law (\ref{eq:A}) 
rewritten as $eV = h\nu$. In facts, he believed the UV radiation 
to arise owing to jumps between discrete energy levels, which 
he deemed to exist and to be gas' property.

Such picture originates on the sole basis of the electric measurements. 
The low-pressure gas is inside the tube, and departures from 
linearity of the mean anodic current as a function of the potentials 
are ascribed to the gas only. If that were true, then it should 
be possible to schematically decompose the circuit into two non-interacting 
parts: a triode that works as a linear amplifier and a variable 
resistance, say an external RCL load having complex impedance. 
The hypothesis is that the low pressure gas in the classical 
electromagnetic theory can be schematised as if it behaved linearly 
in the circuit. That is to say, we attribute to the gas an attenuation 
constant equal to the real coefficient of the impedance, and 
a phase constant equal to the imaginary coefficient \cite{kraus1999}.
If the circuit is supposed to resonate 
at the cascaded RCL mode of the gas, the current maximum (achieved 
at the expenses of the plate battery) corresponds to the load 
driven to resonance.

If the circuit behaved like an amplifier plus a lumped RCL load, 
as measured from the signal detected with a moving-coil mirror 
galvanometer, the input circuit of the amplifying triode should 
provide for a frequency sweep to drive the gas RCL to resonance. 
On the contrary, Einsporn's input supplied static potential differences. 
In order to interpret the experiment according to Einsporn, we 
should substitute the amplifier with a fluorescence spectrometer.

Analysing Goucher's Fig. 3 (our Fig. \ref{fig_3} above), we just saw that a tube devoid 
of gas may perform as a generator at well chosen operating conditions. 
If the frequency of the generator were a linear function of the 
grid potential, then there would be no need to project VCOs (voltage-controlled 
oscillators.) A simple potential-to-frequency conversion formula 
of the kind of law (\ref{eq:A}) would already fit the macroscopic triode 
object. In facts, characteristic curves are already non-linear 
at low frequencies, and at high frequencies the lumped parameter 
wiring diagrams fail altogether.

Fortunately, valves of the kind used by Einsporn, containing 
low pressure mercury vapours, are not really uncommon products. 
They have been implemented in various ways during the last decades. 
They are known in the literature with the name of tiratron. They 
are used as switches in circuits, as alternatives to the ignitron \cite{easman1949}.
 When the valve is on, 
it tends to radiate like a resonator rather than like a linear 
antenna. According to the present interpretation of the Maxwell-Hertz 
theory, the metal making up the body of the cathode is the matter 
being directly electrically strained. It becomes hot. It becomes 
saturated by the power it cannot convert into emission, and radiates 
everything else. After a transient, the metallic surface of the 
catode is \textit{electrically} coupled to anything that can resonate 
in the valve, in particular to the mercury vapours.

Visually, a bluish emission surrounds the cathode, and all other 
luminescence effects which are more easily spectroscopically 
analysed, take place in the tube \cite{seeling1923}.
 Truly, we have included the globally 
non-linear performance of the valve in the term ``saturation''. 
Nevertheless, that doesn't mean one cannot measure ``intermodulation 
products'' in lower frequency ranges by means of a suitable analyser. 
Nor does it mean that linear waves theory cannot be profitably 
used to interpret those measurements. The only prohibition involves 
the kind of spectral analysis implied by law (\ref{eq:A}). Such prohibition 
is prominently related to the paradoxes arising from the subdivision 
of the continuum into discrete parts, and it is purely of mathematical 
nature.

The second series of measures in Einsporn's paper was intended 
for comparison to the new ionisation potentials that Goucher 
had been measuring meanwhile. They have been carried out on the 
tetrod with corrections for taking into account the photoelectric 
effect. As stated in theoretical prescriptions, a strongly negative 
cutoff potential has been applied to the plate, or, with the 
aforementioned corrections, to the screen-grid.

The current vs. potential curve becomes similar to that of a 
diode, exhibiting moderate climbing at the potentials where ionisation 
is expected. That curve indicates it's impossible for the triode 
in those operating conditions to oscillate (off-condition). It 
does not mean that its behaviour becomes linear.

To summarise, the original Franck-Hertz experiment, rather than 
quantitatively confirming a simple law, suggests that physical 
transduction of electrical, thermal or acoustical quantities in radiative 
phenomena is a non-linear performance.

Not that the alternative suggested here, to describe the electric 
interaction between a biased triod and the gas therein as a modulation 
of electromagnetic radiation due to electrical coupling, is easier 
or more convenient to deal with. Nevertheless, if the cavity 
inside triodes could be made selective enough, as wished by many 
of the authors mentioned above, then the modulation could be 
simplified by applying sum and product trigonometric formulas, 
and some additional insight into frequency relationships could 
be gained by comparison with experiments.

\section*{Conclusions}

On reinterpreting those fundamental experiments, the weakness 
of the current concepts about telecommunication theory seems to be 
that it is not based on a signal theory, by which we mean an 
enhanced interpretation of Maxwell's equations. Rather, it's based 
on many communications and information theories, accompanied 
with Lorentz's interpretation of electromagnetic fields.

We have reported the mentioned paper by H. Hertz on the Glimmlicht. 
We have interpreted it with the hindsight given by current advancements 
in the telecommunication field. H. Hertz's experiment appears 
to adhere closely to that interpretation based on Maxwell's equations. 
The reason is precisely that the Author perceived the validation 
of Maxwell's equations as a task separate from, and prevalent 
on, the explanation of specific transduction mechanisms. That 
approach was a good start for founding a detected-signal theory. 
However, the pivotal beliefs around which the research was wandering 
at the time are the same for all the Authors we mentioned. They 
concern the discrete nature of matter and the dynamical model 
of electromagnetism.

Are those beliefs still current?

In practice, the molecular model of the matter has always been 
agreed upon universally. Except for E. Mach, who criticised it 
basing on his own results in optoacoustic experiments. He didn't 
accept it because he couldn't figure out just how statistical 
movements of particles would propagate as nearly monochromatic 
waves \cite{mach1873} or elicit them.

After H. Hertz's time, the relationships among the world out 
there, the sensitive experience, and its interpretation, have 
been the subjects of much writings and discussions. Today, a 
few people believe that Bohr's atom is a faithful description 
of something that really exists in the physical world. Most consider 
it a useful intuitive guide. A model, that is.

In this respect, atomism in general may be thought of as a linear 
model of those relationships with the external world. Working 
on the resulting representation, we can take advantage of its 
simplifications. However, we'll also be hindered by its limitations: 
no further approximation can be easily obtained by introducing 
small perturbations. In facts, it is not possible to use a linear 
model to represent an interaction, be it dynamic, energetic or 
whatever.

As far as a purely mechanical explanation of electromagnetism 
is concerned, we know that Maxwell himself discarded it. After 
the success story of telecommunications, if we accept that electromagnetism 
just explains itself, we don't see any special reason to model 
the current by analogy with convective movement. Moreover, as 
we avoided the previous analogy, we should have no reason to 
bind the charge to a mechanical entity either.

In order to sketch a detected-signal theory, let's reconsider 
the diode for a moment. By varying the potential difference between 
the electrodes, the valve, as it rectifies the current, allows 
its detection. If we don't interpret that current as a flow of 
particles, we may consider it as just the way a diode detects 
varying potentials. In radiotechniques, people talk of rectification 
of electromagnetic oscillations: of a signal, that is.

\end{document}